# Dense Error-Correcting Codes in the Lee Metric

Tuvi Etzion
Department of Computer Science
Technion-Israel Institute of Technology
Haifa 32000, Israel
Email: etzion@cs.technion.ac.il

Alexander Vardy
Department of Electrical
and Computer Engineering
University California, San Diego
La Jolla, CA 92093, USA
Email: avardy@ece.ucsd.edu

Eitan Yaakobi
Department of Electrical
and Computer Engineering
University California, San Diego
La Jolla, CA 92093, USA
Email: eyaakobi@ucsd.edu

*Abstract*—Several new applications and a number of new mathematical techniques have increased the research on error-correcting codes in the Lee metric in the last decade. In this work we consider several coding problems and constructions of error-correcting codes in the Lee metric. First, we consider constructions of dense error-correcting codes in relatively small dimensions over small alphabets. The second problem we solve is construction of diametric perfect codes with minimum distance four. We will construct such codes over various lengths and alphabet sizes. The third problem is to transfer an $n$-dimensional Lee sphere with large radius into a shape, with the same volume, located in a relatively small box. Hadamard matrices play an essential role in the solutions for all three problems. A construction of codes based on Hadamard matrices will start our discussion. These codes approach the sphere packing bound for very high rate range and appear to be the best known codes over some sets of parameters.

## I. INTRODUCTION

The Lee metric was introduced in [1], [2] for transmission of signals taken from GF($p$) over certain noisy channels. It was generalized for $\mathbb{Z}_m$ in [3]. The Lee distance $d_L(X,Y)$ between two words $X = (x_1, x_2, \ldots, x_n)$, $Y = (y_1, y_2, \ldots, y_n) \in \mathbb{Z}_m^n$ is given by $\Sigma_{i=1}^n \min\{x_i - y_i(\bmod\ m), y_i - x_i(\bmod\ m)\}$. A related metric, the Manhattan metric, is defined for alphabet letters taken from the integers. For two words $X = (x_1, x_2, \ldots, x_n), Y = (y_1, y_2, \ldots, y_n) \in \mathbb{Z}^n$ the Manhattan distance between $X$ and $Y$ is defined as $d_M(X,Y) = \Sigma_{i=1}^n |x_i - y_i|$. A code $\mathbb{C}$ in either metric has minimum distance $d$ if for each two distinct codewords $c_1, c_2 \in \mathbb{C}$ we have $d(c_1, c_2) \geq d$, where $d(\cdot, \cdot)$ stands for either the Lee distance or the Manhattan distance. Linear codes are usually the codes which can be handled more effectively and hence we will assume throughout this paper that all codes are linear.

A linear code in $\mathbb{Z}^n$ is an integer lattice. A *lattice* $\Lambda$ is a discrete, additive subgroup of the real $n$-space $\mathbb{R}^n$. W.l.o.g., we can assume that

$$\Lambda = \{u_1 v_1 + u_2 v_2 + \cdots + u_n v_n \ : \ u_1, u_2, \cdots, u_n \in \mathbb{Z}\} \quad (1)$$

where $\{v_1, v_2, \ldots, v_n\}$ is a set of linearly independent vectors in $\mathbb{R}^n$. A lattice $\Lambda$ defined by (1) is a sublattice of $\mathbb{Z}^n$ if and only if $\{v_1, v_2, \ldots, v_n\} \subset \mathbb{Z}^n$. We will be interested solely in sublattices of $\mathbb{Z}^n$. The vectors $v_1, v_2, \ldots, v_n$ are called *basis* for $\Lambda \subseteq \mathbb{Z}^n$, and the $n \times n$ matrix

$$\mathbf{G} = \begin{bmatrix} v_{11} & v_{12} & \ldots & v_{1n} \\ v_{21} & v_{22} & \ldots & v_{2n} \\ \vdots & \vdots & \ddots & \vdots \\ v_{n1} & v_{n2} & \ldots & v_{nn} \end{bmatrix}$$

having these vectors as its rows is said to be a *generator matrix* for $\Lambda$. The lattice with the generator matrix $\mathbf{G}$ is denoted by $\Lambda(\mathbf{G})$.

*Remark 1:* There are other ways to describe a linear code in the Manhattan metric and the Lee metric. The traditional way of using a parity-check matrix can be also used [4], [5]. But, in our discussion, the lattice representation is the most convenient.

The *volume* of a lattice $\Lambda$, denoted $V(\Lambda)$, is inversely proportional to the number of lattice points per unit volume. More precisely, $V(\Lambda)$ may be defined as the volume of the *fundamental parallelogram* $\Pi(\Lambda)$, which is given by

$$\Pi(\Lambda) \stackrel{\text{def}}{=} \{\xi_1 v_1 + \xi_2 v_2 + \cdots + \xi_n v_n \ : \ 0 \leq \xi_i < 1, \ 1 \leq i \leq n\}$$

There is a simple expression for the volume of $\Lambda$, namely, $V(\Lambda) = |\det \mathbf{G}|$.

We say that $\Lambda$ induces a *lattice tiling* of a shape $\mathcal{S}$ if disjoint copies of $\mathcal{S}$ placed on the lattice points on a given specific point in $\mathcal{S}$ form a tiling of $\mathbb{Z}^n$.

Codes in $\mathbb{Z}^n$ generated by a lattice are periodic. We say that the code $\mathbb{C}$ has period $(m_1, m_2, \ldots, m_n) \in \mathbb{Z}^n$ if for each $i, 1 \leq i \leq n$, the word $(x_1, x_2, \ldots, x_n) \in \mathbb{Z}^n$ is a codeword if and only if $(x_1, \ldots, x_{i-1}, x_i + m_i, x_{i+1}, \ldots, x_n) \in \mathbb{C}$. Let $m$ be the least common multiplier of the integers $m_1, m_2, \ldots, m_n$. The code $\mathbb{C}$ has also period $(m, m, \ldots, m)$ and the code $\mathbb{C}$ can be reduced to a code $\mathbb{C}'$ in the Lee metric over the alphabet $\mathbb{Z}_m$ with the same minimum distance as $\mathbb{C}$. The parameters of such code will be given by $(n, d, v, q)$, where $n$ is the length of the code, $d$ is its minimum distance, $v$ is the volume of the related lattice ($\log_q v$ is the redundancy of the code), and $q$ is the alphabet size.

The research on codes with the Manhattan metric is not extensive. It is mostly concerned with the existence and nonexistence of perfect codes [3], [6], [7], [8]. Nevertheless, all codes defined in the Lee metric over some finite alphabet can be extended to codes in the Manhattan metric over the integers. The literature on codes in the Lee metric is very extensive, e.g. [4], [9], [10], [11], [12], [13]. The interest in Lee codes has been increased in the last decade due to many new and diverse applications of these codes. Some examples are constrained and partial-response channels [4], interleaving schemes [14], multidimensional burst-error-correction [15], and error-correction for flash memories [16]. The increased interest is also due to new attempts to settle the existence question of perfect codes in these metrics [8].

The rest of this paper is organized as follows. In Section II we describe two constructions of codes in the Manhattan metric based on Hadamard matrices. The codes constructed can be reduced to codes in the Lee metric over relatively small alphabets. These codes will be very useful in solving our problems in the follow up sections. These codes also approach the sphere packing bound for very high rate range and appear to be the best known codes over infinitely many sets of parameters. In Section III we will consider codes of small length $n$, large minimum distance, over a relatively small alphabet size. Constructions of such codes are closely related to the problem on the highest packing density of cross-polytope over an $n$-dimensional space. In Section IV we will present codes which attain the code-anticode bound, a generalization of the sphere packing bound. While perfect codes with minimum distance three are well known, optimal such codes with minimum distance four are new. We also raise the question of maximum anticodes in all dimensions and diameters. In Section V we present a transformation of $\mathbb{R}^n$ into $\mathbb{R}^n$ such that a cross-polytope is transferred to a shape with a similar volume which can be inscribed in a relatively small box. The transformation is modified to transform only $\mathbb{Z}^n$ into $\mathbb{Z}^n$ and a Lee sphere is transferred to a shape with a similar volume which can be inscribed in a relatively small box. We summarize in Section VI.

## II. HADAMARD MATRICES CODES

In this section we consider constructions of codes based on Hadamard matrices. A Hadamard matrix $\mathcal{H}$ of order $n$ is an $n \times n$ matrix of +1's and -1's such that $\mathcal{H} \cdot \mathcal{H}^t = nI_n$, where $I_n$ is the identity matrix of order $n$. It is well known [17] that if an Hadamard matrix of order $n$ exists then $n = 1, 2$ or $n \equiv 0 (\bmod\ 4)$. There are many constructions for Hadamard matrices and they are known to exist for many values in this range. The first value in this range for which no Hadamard matrix is known yet is $n = 668$ [18]. Each Hadamard matrix can be written in a normal form such that its first row and its first column consist only of +1's. We will now consider only Hadamard matrices written in normal form. Each Hadamard matrix of order $n$ can be considered as a generator matrix of a code over the integers in the Manhattan metric. For such a code we can prove the following theorem.

*Theorem 1:* Let $\mathcal{H}$ be a Hadamard matrix of order $n$. If $\mathcal{H}$ is considered as a generator matrix for a code $\mathbb{C}$ of length $n$ in the Manhattan distance then:
- The minimum distance of $\mathbb{C}$ is $n$.
- The volume of the lattice $\Lambda(\mathcal{H})$ is $n^{\frac{n}{2}}$.
- $\mathbb{C}$ can be reduced to a code $\mathbb{C}'$ of length $n$ in the Lee metric over the alphabet $\mathbb{Z}_n$.

The next construction is based on Hadamard matrices obtained by the doubling construction for length that is a power of two. This code is also based on the first order Reed-Muller code. Hadamard matrix of order 4 is given by

$$\begin{bmatrix} +1 & +1 & +1 & +1 \\ +1 & -1 & +1 & -1 \\ +1 & +1 & -1 & -1 \\ +1 & -1 & -1 & +1 \end{bmatrix}.$$

It generates the same code as the generator matrix

$$\begin{bmatrix} 1 & 1 & 1 & 1 \\ 0 & 2 & 0 & 2 \\ 0 & 0 & 2 & 2 \\ 0 & 0 & 0 & 4 \end{bmatrix}.$$

Reducing the entries into *zeroes* and *ones*, we use the following recursive construction of matrices.

$$H_2 = \begin{bmatrix} 1 & 1 & 1 & 1 \\ 0 & 1 & 0 & 1 \\ 0 & 0 & 1 & 1 \\ 0 & 0 & 0 & 1 \end{bmatrix},$$

$$H_{i+1} = \begin{bmatrix} H_i & H_i \\ 0 & H_i \end{bmatrix}.$$

Let $G(i,j)$, be the $2^i \times 2^i$ matrix constructed from $H_i$ as follows. Let $2^\ell$ be the sum of elements in row $s$ of $H_i$. If $2^\ell \geq 2^j$ then row $s$ of $G(i,j)$ will be the same as row $s$ of $H_i$. If $2^\ell < 2^j$ then row $s$ of $G(i,j)$ will be row $s$ of $H_i$ multiplied by $2^{j-\ell}$.

*Theorem 2:* If $G(i,j)$ is considered as a generator matrix for a code $\mathbb{C}$ of length $2^i$ in the Manhattan distance then:
- The minimum distance of $\mathbb{C}$ is $2^j$.
- The volume of the lattice $\Lambda(G(i,j))$ is $\Pi_{r=0}^{\min\{i,j\}} 2^{(j-r)\binom{i}{r}}$.
- $\mathbb{C}$ can be reduced to a code $\mathbb{C}'$ of length $2^i$ in the Lee metric over the alphabet $\mathbb{Z}_{2^j}$.

The codes obtained from $G(i,j)$ approach the sphere packing bound for very high rate range as the codes of Roth and Siegel [4].

The construction of the generator matrices $G(i,j)$ yields codes of length $2^i$, $i \geq 2$, with minimum distance $2^j$, $j \geq 2$. In a similar way we can construct codes of length $\alpha 2^i$, $i \geq 2$, $\alpha$ an odd integer, with minimum distance $\alpha 2^j$, $j \geq 2$. The construction makes use of a code of length $4 \cdot \alpha$ obtained from a Hadamard matrix of order $4 \cdot \alpha$. Similar results to the ones in Theorem 2 are obtained.

## III. DENSE CODES WITH LARGE MINIMUM DISTANCE

An $n$-dimensional cross-polytope is the shape defined by the equation

$$\Sigma_{i=1}^n |x_i| \leq 1 .$$

The packing density of a lattice $\Lambda$ which defines a packing for a shape $\mathcal{S}$ is defined by $\frac{|\mathcal{S}|}{V(\Lambda)}$. What is the densest lattice packing of an $n$-dimensional cross-polytope? For $n = 2$ the required packing density is 1, and it is implied

by the related perfect codes in the Lee metric [3], [14]. For $n = 3$, the packing density proved to be $\frac{18}{19}$ and it was proved by Minkowski [19]. For $n > 3$ only a few lower bounds are known on the packing density of the $n$-dimensional cross-polytope, $\theta(n)$. A short survey on the known results is given in [20]. In [21] it is proved that $\theta(4) \geq \frac{512}{621} = 0.82447\ldots$. Cools [20] showed that $\theta(5) \geq \frac{1600}{2343} = 0.68288\ldots$. Cools and Govaert [22] proved the bound $\theta(6) \geq \frac{38416}{71595} = 0.53657\ldots$. For higher dimensions no result is known. But, also for dimensions 4, 5, and 6, the given lattices are not practical from coding theory point of view.

An $n$-*dimensional Lee sphere* with radius $R$ is the shape $S_{n,R}$ in $\mathbb{Z}^n$ such that $(x_1, x_2, ..., x_n) \in S_{n,R}$ if and only if $\sum_{i=1}^{n} |x_i - y_i| \leq R$, i.e., it consists of all points in $\mathbb{Z}^n$ whose distance from a given point $(y_1, y_2, ..., y_n) \in \mathbb{Z}^n$ is at most $R$. The size of $S_{n,R}$ is well known [3]:

$$|S_{n,R}| = \sum_{i=0}^{\min\{n,R\}} 2^i \binom{n}{i}\binom{R}{i} \quad (2)$$

The size of an $n$-dimensional Lee sphere with radius $\frac{d}{2}$ is $\frac{d^n}{n!} + O(d^{n-1})$ when $n$ is fixed and $d \longrightarrow \infty$ (an $n$-dimensional code with even minimum distance $d$ in $\mathbb{Z}^n$ forms a packing for an $n$-dimensional Lee sphere with radius $\frac{d}{2}$).

Our goal in this section is to find lattices with good packing density of the $n$-dimensional cross-polytope, which induce an error-correcting code of length $n$, large minimum distance, over relatively small alphabet size. They will also induce high packing density of the related Lee sphere. We will use four types of constructions:

1) Direct constructions for small dimension.
2) Constructions based on Hadamard matrices.
3) Recursive constructions.
4) Puncturing.

We start with the recursive construction. Let $G_i$, $i = 1, 2$, be a generator matrix for an $(n_i, d_i, v_i, q_i)$ code, which has $\frac{(d_i)^{n_i}}{n_i! \cdot v_i}$ packing density. Then the generator matrix defined by the direct product $G_1 \times G_2$ is a generator matrix of an $(n_1 n_2, d_1 d_2, (v_1)^{n_2} (v_2)^{n_1}, q_1 q_2)$ code (lattice) with $\frac{(d_1 d_2)^{n_1 n_2}}{(n_1 n_2)! \cdot (v_1)^{n_2} (v_2)^{n_1}}$ packing density.

For direct constructions we will consider $n = 3, 4, 5$ and 7. For $n = 3$ we use the lattice of Minkowski [19] with the generator matrix

$$\frac{d}{6} \begin{bmatrix} 1 & -2 & 3 \\ -2 & 3 & 1 \\ 3 & 1 & -2 \end{bmatrix}$$

to obtain a $(3, d, \frac{19}{108}d^3, 19\frac{d}{3})$ code with packing density $\frac{18}{19}$. For $n = 4$ we use the lattice with the generator matrix

$$\frac{d}{6} \begin{bmatrix} 1 & 0 & 1 & 4 \\ 0 & 1 & 0 & 7 \\ 0 & 0 & 1 & 23 \\ 0 & 0 & 0 & 74 \end{bmatrix}$$

to obtain a $(4, d, \frac{13}{216}d^4, 37\frac{d}{3})$ code with packing density $\frac{9}{13}$. For $n = 5$ we use a $(5, d, \frac{5}{256}d^5, 5d)$ code with packing density $\frac{32}{75}$. For $n = 7$ we use a $(7, d, \frac{7}{2^{12}}d^7, 7d)$ code with packing density $0.1161\ldots$. These codes are discussed in the next section as they are diameter perfect codes with minimum distance 4.

Using the recursive construction for $n = 6$ from $(2, d_2, \frac{(d_2)^2}{2}, d_2)$ code and the $(3, d_3, \frac{19}{108}(d_3)^3, 19\frac{d_3}{3})$ code we obtain a $(6, d, \frac{1}{2}(\frac{19}{216})^2, 19\frac{d}{3})$ code, where $d = d_2 d_3$, with packing density $\frac{2}{5}(\frac{18}{19})^2 = 0.359\ldots$. Similarly, codes for $n = 8, 9,$ and 10, are constructed.

Codes with lower packing density, but with smaller alphabet size, are obtained with the generator matrices $G(i, j)$ or other generator matrices derived from a Hadamard matrices.

Finally, we consider puncturing. Assume we have an $(n+1, d, v, q)$ code with packing density $\theta$. Assume further that the first entry in the diagonal of the generator matrix is a *one* (if it is a larger entry a better result is obtained). We remove the first row and column of the generator matrix to obtain an $(n, d, v, q)$ code. Puncturing is effective to obtain a code of length $n$, where $n$ is a large odd integer and has only a trivial factorization. The first such case is $n = 11$.

## IV. MAXIMUM DENSITY ERROR-CORRECTING CODES

In this section we will consider error-correcting codes with maximum density in both metrics. For length which is a power of two the alphabet will be $\mathbb{Z}_4$. We start by considering perfect codes and after that we will consider diameter perfect codes. In between we will consider the problem of maximum anticodes in the Manhattan metric.

### A. Perfect error-correcting codes

For any given metric in coding theory it is quite popular and interesting to search for perfect codes. A code $\mathbb{C}$ is called $e$-perfect if for any word $x$ there exists exactly one codeword $c \in \mathbb{C}$ such that $d(x, c) \leq e$. The following theorem is known as the sphere packing bound:

*Theorem 3:* For a code $\mathbb{C} \subseteq \mathbb{Z}_q^n$ with minimum distance $2R + 1$ we have $|\mathbb{C}| \cdot |S_{n,R}| \leq q^n$.

A code $\mathbb{C}$ which attains the bound of Theorem 3 is an $R$-perfect code. If the code $\mathbb{C}$ is over $\mathbb{Z}^n$ and the code $\mathbb{C}$ is $R$-perfect then the spheres with radius $R$ around the codewords form a tiling of $\mathbb{Z}^n$. Instead of Theorem 3 we have the following theorem.

*Theorem 4:* For a code $\mathbb{C} \subseteq \mathbb{Z}^n$ with minimum distance $2R + 1$, whose codewords are the points of a lattice $\Lambda$ we have $V(\Lambda) \geq |S_{n,R}|$.

$R$-perfect codes are known for $n = 2$ and any given $R$ for alphabet size divisible by the size of the sphere. For larger $n$, $R$-perfect codes are known for $R = 1$ and alphabet size divisible by $2n + 1$. These codes were described by Golomb and Welch [3]. The existence of other $R$-perfect codes was discussed in many papers [3], [6], [7], [8].

### B. Diameter Perfect Codes

The concept of perfect code is applied to the case when the minimum distance of the code is an odd integer. If the minimum distance of the code is an even integer then there cannot be any perfect code since for two codewords $c_1, c_2 \in \mathbb{C}$ such that $d(c_1, c_2) = 2\delta$ there exists a word $x$ such that

$d(x, c_1) = \delta$ and $d(x, c_2) = \delta$. For this case another concept is used, diameter perfect code, as was defined in [23]. This concept is based on the code-anticode bound presented by Delsarte [24]. An anticode $\mathcal{A}$ of diameter $D$ is a subset of words such that $d(x, y) \leq D$ for all $x, y \in \mathcal{A}$.

*Theorem 5:* If a code $\mathbb{C}$ of length $n$ over $\mathbb{Z}_q$ has minimum distance $d$ and in the anticode $\mathcal{A}$ of length $n$ over $\mathbb{Z}_q$ the maximum distance is $d - 1$ then $|\mathbb{C}| \cdot |\mathcal{A}| \leq q^n$.

*Theorem 6:* Let $\Lambda$ be a lattice which forms a code $\mathbb{C} \subseteq \mathbb{Z}^n$ with minimum distance $d$. Then $V(\Lambda)$ is at least the size of any anticode of length $n$ with maximum distance $d - 1$.

Codes which attain the bounds of Theorems 5 and 6 with equality are called diameter perfect codes.

*C. Maximum anticodes*

What is the size of the largest anticode with Diameter $D$ in $\mathbb{Z}^n$, $n \geq 2$. If $D = 2R$ we conjecture that this anticode is a sphere with radius $R$, $S_{n,R}$. For odd $D$ we define anticodes with diameter $D = 2R+1$ as follows. For $R = 0$, let $S'_{n,0}$ be a shape which consists of two adjacent points of $\mathbb{Z}^n$, where two points are adjacent if the distance between them is one. $S'_{n,R+1}$ is defined by adding to $S'_{n,R}$ all the points which are adjacent to at least one of its points.

*Lemma 1:* $|S'_{n,R}| = \sum_{i=0}^{\min\{n-1,R\}} 2^{i+1} \binom{n-1}{i} \binom{R+1}{i+1}$.

*Lemma 2:*

- $|S_{n,R}| = |S_{n-1,R}| + |S'_{n,R-1}|$.
- $|S'_{n,R}| = |S_{n-1,R}| + |S_{n,R}|$.
- $|S_{n,0}| = 1$, and for $R > 0$, $|S_{1,R}| = 2R + 1$ and $|S'_{1,R}| = 2R + 2$.

We conjecture that $S_{n,R}$ is the maximum anticode of length $n$ and diameter $2R$ and $S'_{n,R}$ is the maximum anticode of length $n$ and diameter $2R+1$. We have been able to prove this conjecture for some values of $R$. The size of a maximum anticode when the anticode is linear was determined in [25]. Of course, usually these anticodes are not linear. In other metrics the problem of finding the maximum anticodes was extensively studied (see [23] for some discussion). For example, the Hamming space was considered in [26].

*D. Constructions of Codes*

Perfect codes with minimum distance three were constructed in [3] and from lack of space we omit discussion on these codes. We will discuss now diameter perfect codes with minimum distance four. We will consider two constructions.

Consider the lattice $\Lambda_n$ defined by the following generator matrix

$$G_n = \begin{bmatrix} A_n & B_n \\ C_n & D_n \end{bmatrix}$$

where $A_n = I_{n-1}$, $B_n$ is the $(n-1) \times 1$ matrix for which $B_n^T = [3\ 5\ \cdots\ 2n-1]$, $C_n$ is an $1 \times (n-1)$ allzero matrix, and $D_n = [4n]$.

*Example 1:* For $n = 6$, $G_6$ has the form

$$G_6 = \begin{bmatrix} 1 & 0 & 0 & 0 & 0 & 3 \\ 0 & 1 & 0 & 0 & 0 & 5 \\ 0 & 0 & 1 & 0 & 0 & 7 \\ 0 & 0 & 0 & 1 & 0 & 9 \\ 0 & 0 & 0 & 0 & 1 & 11 \\ 0 & 0 & 0 & 0 & 0 & 24 \end{bmatrix}$$

*Theorem 7:* The code $\mathbb{C}$ whose generator matrix is $G_n$ is a diameter perfect code with parameters $(n, 4, 4n, 4n)$.

The well known doubling construction for perfect codes in the Hamming scheme [27] also works here to obtain diameter perfect codes (linear and nonlinear). Theorem 7 is replaced with the following theorem.

*Theorem 8:* For a given positive odd integer $n$, and any integer $i \geq 0$, there exists a diameter perfect code with parameters $(2^i n, 4, 2^{i+2} n, 4n)$.

When $n$ is a power of two we can obtain a diameter perfect code with parameters $(2^i, 4, 2^{i+2}, 4)$, by using $G(i, 2)$ as a generator matrix. Codes with the same parameters were generated by Krotov [28].

## V. Lee sphere Transformation

In multidimensional coding, many techniques work only on multidimensional boxes and do not work on other shapes, e.g. [15], [29], [30]. One solution to overcome this problem when we need to imply our coding problem on a different $n$-dimensional shape $\mathcal{S}$ is to inscribe $\mathcal{S}$ in a relatively small $n$-dimensional box. One of the most important shapes in this context is an $n$-dimensional Lee sphere with radius $R$, $S_{n,R}$. An $n$-dimensional Lee sphere with radius $R$ can be inscribed in an $\underbrace{(2R+1) \times \cdots \times (2R+1)}_{n \text{ times}}$ $n$-dimensional box. In [15] a transformation of $\mathbb{Z}^n$ is given for which $S_{n,R}$ is transformed into a shape inscribed in a box of size $\underbrace{(R+1) \times (R+1) \times \cdots \times (R+1)}_{n-1 \text{ times}} \times (2R+1)$. The gap from the theoretical size of the box is still large since the size of the $n$-dimensional Lee sphere with radius $R$ is $\frac{(2R)^n}{n!} + O(R^{n-1})$ when $n$ is fixed and $R \longrightarrow \infty$. The goal of this section is to close on this gap. The transformation $T : \mathbb{Z}^n \longrightarrow \mathbb{Z}^n$ is clearly a discrete transformation, but for completeness we will consider also the more simple case of a continuous transformation $T : \mathbb{R}^n \to \mathbb{R}^n$.

*A. The Continuous Transformation*

Let $\mathcal{H}$ be a Hadamard matrix of order $n \geq 4$. We define the transformation $T : \mathbb{R}^n \to \mathbb{R}^n$ as follows. For each $\boldsymbol{x} = (x_1, \ldots, x_n) \in \mathbb{R}^n$,

$$T(\boldsymbol{x}) = \frac{\mathcal{H} \cdot \boldsymbol{x}}{\sqrt{n}}. \qquad (3)$$

The following two lemmas which can be easily verified are the key of our discussion.

*Lemma 3:* For all $\boldsymbol{x} = (x_1, \ldots, x_n) \in \mathbb{R}^n$,

$$T(T(\boldsymbol{x})) = \boldsymbol{x}.$$

*Lemma 4:* A Lee sphere with radius $R$ is inscribed after the transformation $T$ inside an $n$-dimensional box of size

$$\frac{2R}{\sqrt{n}} \times \cdots \times \frac{2R}{\sqrt{n}}.$$

Note that since the transformation $T$ is an affine transformation and $\det(\mathcal{H}/\sqrt{n}) = 1$ then it also preserves volume. The following lemma will be useful for a discrete transformation.

*Theorem 9:* Let $\mathbb{C}$ be the list of words which are mapped to points of $\mathbb{Z}^n$ by the transformation $T$ of (3) using a Hadamard matrix $\mathcal{H}$ of order $n$. If $n = d^2$, where $d$ is a positive integer, then $\mathbb{C}$ is an error-correcting code with minimum Lee distance $d$; moreover $\mathbb{C} = T(\mathbb{C})$, where $T(\mathbb{C}) = \{T(c) \ : \ c \in \mathbb{C}\}$.

### B. The Discrete Transformation

For the discrete case we want to modify the transformation $T$ used for the continuous case. Let $d \geq 2$ be a positive integer and $\mathcal{H}$ a Hadamard matrix of order $d^2$. Let $S$ be the set of coset leaders of the code $\mathbb{C}$ defined in Theorem 9. The transformation $T_{d^2} : \mathbb{Z}^{d^2} \to \mathbb{Z}^{d^2}$ is defined as follows: for each $(i_1, \ldots, i_{d^2}) \in \mathbb{Z}^{d^2}$, we write $(i_1, \ldots, i_{d^2}) = (c_1, \ldots, c_{d^2}) + (s_1, \ldots, s_{d^2})$, where $(c_1, \ldots, c_{d^2}) \in \mathbb{C}$ and $(s_1, \ldots, s_{d^2}) \in S$.

$$T_{d^2}(i_1, \ldots, i_{d^2}) = T(c_1, \ldots, c_{d^2}) + (s_1, \ldots, s_{d^2}),$$

where $T$ is defined in (3).

*Theorem 10:* For all $(i_1, \ldots, i_{d^2}) \in \mathbb{Z}^{d^2}$,

$$T_{d^2}(T_{d^2}(i_1, \ldots, i_{d^2})) = (i_1, \ldots, i_{d^2}).$$

Let $\rho$ be the covering radius of the code $\mathbb{C}$ defined in Theorem 9.

*Lemma 5:* A Lee sphere with radius $R$ is inscribed after the transformation $T_{d^2}$ inside an $n$-dimensional box of size

$$\left(2\left\lceil\frac{R+\rho}{d}\right\rceil + 2\rho + 1\right) \times \cdots \times \left(2\left\lceil\frac{R+\rho}{d}\right\rceil + 2\rho + 1\right).$$

## VI. CONCLUSION

Several coding problems in the related Manhattan metric and Lee metric were discussed in this paper. We considered dense error-correcting codes in the Lee scheme with a relatively small size alphabet. We considered dense packing of cross-polytopes in arbitrary dimension. Perfect codes (and also perfect diameter codes) in the Lee metric were discussed. The problem of maximum anticodes in the Manhattan metric was considered. We examined codes in these metrics which are derived from Hadamard matrices. Finally, a transformation of $\mathbb{Z}^n$ to itself in which a Lee sphere was transformed into a shape, with the same volume, inscribed in a relatively small box was given.

## ACKNOWLEDGMENT

This work was supported in part by the United States-Israel Binational Science Foundation (BSF), Jerusalem, Israel, under Grant No. 2006097. The work of Eitan Yaakobi was supported in part by the University of California Lab Fees Research Program, Award No. 09-LR-06-118620-SIEP.